\documentclass[prl,twocolumn,showpacs,preprintnumbers,amsmath,amssymb,tightenlines,epsfig]{revtex4}

\usepackage{graphicx}
\usepackage{dcolumn}
\usepackage{bm}

\begin{document}

\title{Bose-Einstein condensate collapse: a comparison between theory and experiment}
\author{C.M. Savage,  N.P. Robins, and J.J. Hope}
\affiliation{Department of Physics and Theoretical Physics, 
Australian National University,
ACT 0200, Australia}

\email{craig.savage@anu.edu.au}

\begin{abstract}
We  solve the Gross-Pitaevskii equation numerically for the collapse induced by a switch from positive to negative scattering lengths. We compare our results with experiments performed at JILA with  Bose-Einstein condensates of $^{85}$Rb, in which the scattering length was controlled using a Feshbach resonance. Building on previous theoretical work we identify quantitative differences between the predictions of mean-field theory and the results of the experiments. Besides the previously reported difference between the predicted and observed critical atom number for collapse, we also find that the predicted collapse times systematically exceed those observed experimentally. Quantum field effects, such as fragmentation, that might account for these discrepancies are discussed.
\end{abstract}

\maketitle


\textit{Introduction.}-- Most experiments on dilute gas Bose-Einstein condensates (BECs) are performed with atoms that have a repulsive two-body interaction. Exceptions are the experiments on $^{7}$Li \cite{Bradley,Gerton} and, more recently, on $^{85}$Rb \cite{Donley,Robertscollapse}. For $^{85}$Rb a Feshbach resonance allows the two-body interaction strength to be tuned over a wide range of attractive and repulsive values. In particular, the scattering length has been rapidly switched from positive (repulsive interaction) to negative (attractive interaction) values, leading to the collapse and subsequent explosion of the condensate. Recently, the large positive scattering lengths attainable in this system have been used to produce atom-molecule condensates \cite{Donleymolecule}.

In the following we report on our modelling of the $^{85}$Rb collapse experiments \cite{Donley}, using the Gross-Pitaevskii (GP) equation for the expectation value of the field operator \cite{Dalfovo,RMP,MeystreAObook,PethickSmith}.  Saito and Ueda \cite{Saitoexplosion} and Adhikari \cite{Adhikariexplosion} have also modelled these experiments by numerical solution of the cylindrically symmetric GP equation.  Saito and Ueda conclude that this describes the collapsing and exploding dynamics at least qualitatively \cite{Saitoexplosion}. Following their  suggestion, we report a more quantitative comparison between the theoretical and experimental results, and find significant differences.

The series of experiments on the collapse and explosion of $^{85}$Rb BECs challenges theoretical models in a number of ways  \cite{Claussen}. A body of theoretical work based on the GP equation  predicts the critical number of atoms $N_{\rm{cr}}$ for collapse to be significantly larger than is observed. The expression for the critical number is
\begin{equation}
N_{\rm{cr}} = k \frac{a_{\rm{ho}}} {|a|} ,
\label{critical number}
\end{equation}
where $a_{\rm{ho}} = \sqrt{\hbar / (m \bar{\omega})}$ is the harmonic oscillator scale length, with $\bar{\omega}$ the geometric mean of the trap frequencies in the three Cartesian directions,  and $a$ the scattering length. Experimentally, $k=0.46 \pm 0.06$  \cite{Robertscollapse}, whereas $k=0.57$ according to various approximate solutions of the GP equation \cite{EleftheriouHuang,Ruprecht}. 

We have confirmed this GP prediction for the specific cylindrically symmetric experimental case  \cite{Robertscollapse} with cylindrically symmetric numerical solutions. We verified these with full  three dimensional numerical solutions, and also confirmed that slight departures from cylindrical symmetry had no effect on the critical number \cite{3dnumerics}. Consequently there is a disagreement at the two standard deviations level, which should be regarded as significant.

We also report a new quantitative discrepancy between the predictions of the GP model and experiment. Under certain conditions, the GP predicted time to the initiation of collapse, $t_{\rm collapse}$, is systematically longer than that observed in the experiments \cite{Donley}.


\textit{The GP model.}-- In the conclusion we will discuss the possiblility that these discrepancies result from quantum field effects beyond the GP approximation. We therefore now derive the GP equation from the quantum field theory. 

The second-quantised Hamiltonian for a dilute gas, in terms of the field operator $\hat{\Psi}(\mathbf{r},t)$, is
\begin{eqnarray}
H &=& \int d \mathbf{r} \hat{\Psi}^\dagger 
H_0 \hat{\Psi} 
\nonumber \\
&& + \frac{1}{2}  \int d \mathbf{r}  d \mathbf{r}' \hat{\Psi}^{\dagger} \hat{\Psi}^{\dagger \prime} 
V(\mathbf{r} - \mathbf{r}') \hat{\Psi}' \hat{\Psi} ,
\label{Hamiltonian}
\end{eqnarray}
where $\hat{\Psi}' = \hat{\Psi}(\mathbf{r}',t)$ and $H_0$ is the single particle Hamiltonian for the kinetic energy and trapping potential 
\begin{equation}
H_0  = -\frac{\hbar^2}{2m} \nabla^2 
+\frac{1}{2} m  ( \omega_x^2 x^2 + \omega_y^2 y^2 +\omega_z^2 z^2 ) ,
\label{H0}
\end{equation}
where $m$ is the atomic mass ($1.41 \times 10^{-25}$ kg for $^{85}$Rb), and $\omega_i$ is the trap frequency along Cartesian axis $i$. In the limit of particles separated by distances much greater than the scattering length $a$ we approximate the two-body potential by a delta function interaction \cite{Dalfovo,RMP,MeystreAObook,PethickSmith}
\begin{equation}
V(\mathbf{r} - \mathbf{r}') = g \delta(\mathbf{r} - \mathbf{r}'), \quad
g = \frac{4 \pi \hbar^2 a}{m} .
\label{delta fn potential}
\end{equation}
The Heisenberg dynamical equation for the field operator is then
\begin{equation}
i \hbar  \frac{\partial}{\partial t}  \hat{\Psi} = 
H_0  \hat{\Psi} + g \hat{\Psi}^\dagger \hat{\Psi}  \hat{\Psi},
\label{heisenberg eq}
\end{equation}
Taking the symmetry-breaking approach we assume that the field expectation value is not zero and define it as the GP wavefunction $\langle \hat{\Psi}(\mathbf{r},t) \rangle = \Phi(\mathbf{r},t)$, normalised to the number of particles $N$
\begin{equation}
N = \int  | \Phi(\mathbf{r},t) |^2 d \mathbf{r} .
\label{GP normalisation}
\end{equation}
Then taking the expectation value of the Heisenberg equation (\ref{heisenberg eq}) gives
\begin{equation}
i \hbar \frac{\partial}{\partial t}  \Phi = H_0  \Phi 
+ g \langle \hat{\Psi}^\dagger \hat{\Psi}  \hat{\Psi} \rangle ,
\label{heisenberg eq expectation value}
\end{equation}
If we assume that the expectation value factorises, as it would, for example, if the system were in an eigenstate of the field operator, 
\begin{equation}
 \langle \hat{\Psi}^\dagger \hat{\Psi}  \hat{\Psi} \rangle = \Phi^* \Phi \Phi,
\label{factorisation}
\end{equation}
then we obtain the GP equation
\begin{equation}
i \hbar \frac{\partial}{\partial t}  \Phi = ( H_0  + g | \Phi |^2  ) \Phi .
\label{GP eq no loss}
\end{equation}
In order to model atom loss due to three-body recombination we add a phenomenological term proportional to the density $| \Phi |^2 $ squared with rate coefficient $K_3 /2$ \cite{Robertsparams,PethickSmith}
\begin{equation}
i \hbar \frac{\partial}{\partial t}  \Phi = 
( H_0  + g | \Phi |^2 - i \frac{\hbar}{2} K_3  | \Phi |^4 ) \Phi .
\label{GP eq}
\end{equation}
We assume one-body and two-body loss are negligible, as was true for the relevant experiments. The number of atoms then decays as
\begin{equation}
\frac{dN}{dt} = -K_3 \int  | \Phi(\mathbf{r},t) |^6 d \mathbf{r} .
\label{loss rate equation}
\end{equation}
%

\begin{figure}
\includegraphics[width=\columnwidth]{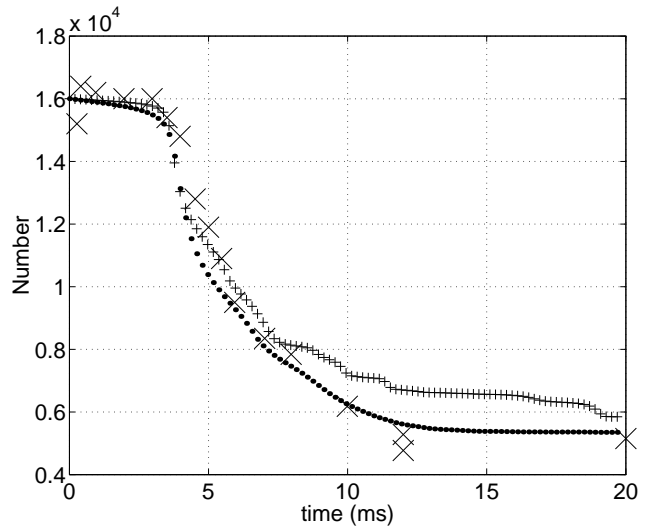}
\caption{Experimental and numerical results for the number of atoms $N$ versus time after a switch from $a = +7a_0$ to $a = -30a_0$. The experimental points ($\times$) are from Fig.~1b of Donley \textit{et al.}  \cite{Donley}. The numerical results are for $K_3 = 190 \times 10^{-28}$ cm$^6$s$^{-1}$ (filled circles) and for $K_3 = 78 \times 10^{-28}$ cm$^6$s$^{-1}$ (+). Other parameters are as given in the experimental paper  \cite{Donley}:  $N_0 =16,000$, radial frequency $\omega_r = 2\pi \times 17.5$ Hz, axial frequency $\omega_z  = 2\pi \times 6.8$ Hz.}
\label{fig:Donley1b}
\end{figure}

\textit{GP Results.}--
As an example of the ability of the GP equation to correctly model the $^{85}$Rb \cite{Donley} experiments we present Fig.~\ref{fig:Donley1b}. It is the result of a numerical solution of the (two dimensional) cylindrically symmetric GP equation for $\tilde{\Phi}(r,z)$
\begin{eqnarray}
i \hbar \frac{\partial}{\partial t}  \tilde{\Phi} &=& 
-\frac{\hbar^2}{2m} (\partial^2_r + r^{-1}\partial_r +\partial^2_z)  \tilde{\Phi} 
\nonumber \\ 
&& +\frac{1}{2} m  ( \omega_r^2 r^2  +\omega_z^2 z^2 )  \tilde{\Phi}
+g | \tilde{\Phi} |^2 \tilde{\Phi} 
\nonumber \\ 
&& - i \frac{\hbar}{2} K_3  | \tilde{\Phi} |^4  \tilde{\Phi}  .
\label{cylindrical GP}
\end{eqnarray}
Parameters are the same those of Fig.~1b of Donley \textit{et al.}~\cite{Donley}. Specifically, the ground state of the GP equation for $a = +7a_0$ was switched in 1 ms to $a = -30a_0$, where $a_0=0.0529$ nm is the Bohr radius. For the three-body recombination rate coefficient $K_3 = 190 \times 10^{-28}$ cm$^6$s$^{-1}$ the agreement with the experimental results is good. However it should be noted that the experimental points are the ``remnant'' atom number, while the numerical points are the total atom number, which overestimates the remnant atom number. A smaller value of $K_3$ agrees better with the earlier points, while overestimating the final atom number. The precise value of $K_3$ has little effect on the conclusions of this paper, which concern the initiation of collapse.

These results agree with those reported by Saito and Ueda \cite{Saitoexplosion} and Adhikari \cite{Adhikariexplosion}. However the former authors used a much smaller value of  the three-body recombination rate coefficient $K_3= 2 \times 10^{-28}$ cm$^6$s$^{-1}$. This produces the collapses and revivals in condensate size that were observed in their simulations. These only become important for $K_3$ less than about $10^{-26}$ cm$^6$s$^{-1}$. Adhikari \cite{Adhikariexplosion} used the much larger value $K_3= 13 \times 10^{-25}$ cm$^6$s$^{-1}$. Since three-body recombination is responsible for the atom loss, it is remarkable that such a wide range of coefficients reproduces the experimental results.

The three-body recombination rate coefficient $K_3$ is expected to vary strongly near the Feshbach resonance \cite{Esry}. Experimental determination of $K_3$ is difficult due to the low densities of $^{85}$Rb condensates. Upper bounds have been estimated to be $5 \times 10^{-25}$ cm$^6$s$^{-1}$, dropping to $10^{-26}$ cm$^6$s$^{-1}$ nearer the Feshbach resonance \cite{Robertsparams}.

The cylindrically symmetric numerical simulations were performed on a $512 \times 512$ grid, 35.64 $\mu$m long in the axial ($z$) direction and with the radial coordinate extending to 11.88 $ \mu$m. The corresponding spatial grid spacings were therefore 0.07 $\mu$m and 0.023 $\mu$m. The time steps were $2.34$ ns.  All simulations were performed on a multiprocessor machine \cite{APAC}, using up to 32 processors, and the RK4IP algorithm developed by the BEC theory group of R. Ballagh at the University of Otago \cite{RK4IP}. This is a pseudo-spectral method with a Runge-Kutta time step. The cylindrically symmetric and full three-dimensional codes are independent and were cross checked. Grid spacings and time steps were varied to ensure convergence. Overall the results were found to be quite robust.  As another test, we solved the GP equation for a half a radial period after the quenching of the collapse.  As was observed experimentally, the condensate refocussed onto the axis, due to the oscillation in the harmonic radial potential. All this, together with the agreement of our results with those of Saito and Ueda \cite{Saitoexplosion} and Adhikari \cite{Adhikariexplosion}, gives us confidence in their accuracy. Following Adhikari \cite{Adhikariexplosion}, the initial condition for Fig.~\ref{fig:Donley1b} was generated by adiabatically expanding the harmonic oscillator initial state $a=0$ to $a = +7a_0$ over 444 ms.

\begin{figure}
\includegraphics[width=\columnwidth]{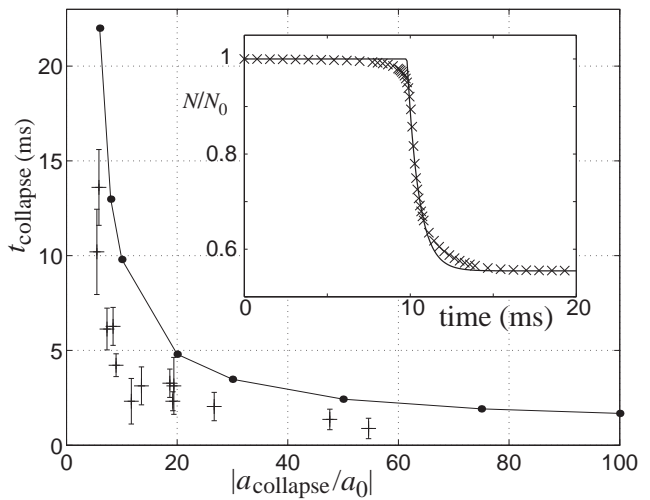}
\caption{Experimental and numerical results for the collapse time $t_{\rm{collapse}}$ versus scattering length $a_{\rm{collapse}}$ after a switch from $a = 0$ to $a_{\rm{collapse}}$. The experimental points (+) and their error bars are from Fig.~2 of Donley \textit{et al.}  \cite{Donley}. The numerical results (filled circles) are for $K_3 = 190 \times 10^{-28}$ cm$^6$s$^{-1}$ . Other parameters are as given in the Fig.~\ref{fig:Donley1b} caption, except:  $N_0 =6,000$.
Inset: Example of the fitting procedure used to determine the collapse times. Shown is a fit of the functional form Eq.(\ref{fit function}) (solid line) to the GP simulation ($\times$) for $a = -10a_0$. The fit parameters here are $t_{\rm{collapse}} = 9.8$ ms, $\tau_{\rm{decay}} = 0.7$ ms, and $N_{\rm{f}}/N_{0}=0.5544$. }
\label{fig:Donley 2}
\end{figure}

Figure~\ref{fig:Donley 2} presents our calculations of the collapse times $t_{\rm{collapse}}$ for the conditions of Fig.~2 of Donley \textit{et al.}  \cite{Donley}. The collapse times were determined by visually fitting plots of atom number versus time to the functional form
\begin{equation}
N = (N_0 -N_{\rm{f}})\exp[-(t-t_{\rm{collapse}})/\tau_{\rm{decay}}]  +N_{\rm{f}},
\label{fit function}
\end{equation}
where $N_{\rm{f}}$ is the long time atom number. An example is given in the inset to Fig.~\ref{fig:Donley 2}.  We have also plotted the experimental results reported in Fig.~2 of Donley \textit{et al.}  \cite{Donley}, and find a small, but significant, systematic disagreement with the GP results. Although the reported errors in the experimental collapse times are large, the GP values for $t_{\rm{collapse}}$ are consistently longer than the experimental ones. This is surprising as the GP model is expected to be valid for the low densities preceding the collapse. If it were to fail, it would be expected to do so at the high densities generated subsequently. Nevertheless, the disagreement is not unprecedented since, as we discussed earlier, the GP model also overestimates the critical number for collapse. 

The estimates of $t_{\rm{collapse}}$ by Saito and Ueda \cite{Saitoexplosion} (their Fig.~3) are between five percent (low $a$) and ten percent (high $a$) smaller than ours. This is consistent with the smaller three-body recombination rate coefficient $K_3$ they used. However their results are still significantly longer than the experimentally measured times. 

We have confirmed these cylindrically symmetric simulations by performing full three-dimensional simulations.  In particular we broke the cylindrical symmetry by using trap frequencies of $17.24 \times 17.47 \times 6.80$ Hz \cite{Robertscollapse}.

We were unable to substantially improve the agreement by either changing the initial condition to reflect the experimental uncertainty of $a= \pm 2 a_0$, or by varying the three-body recombination rate coefficient. This suggests that some of the physics determining the collapse time is not captured by our GP model.

%

\textit{Discussion.}--
Both the collapse time and critical number discrepancies could be resolved by using a scattering length in the GP model larger in magnitude than the experimental value. This would reduce the collapse time and decrease the critical number, as required. The required increases in the scattering length magnitudes vary, ranging from a factor of $0.57/0.46 = 1.2$ for the critical number, up to a factor of about two for the collapse times for large $a_{\rm{collapse}}$. However, the scattering length is experimentally well calibrated \cite{Robertscalibrate}, so any such change would reflect a deficiency of our GP model.

One possible origin of the discrepancy is the effect of thermal non-condensed atoms. Because of the quantum statistics of collisions betwen bosons, the scattering length between a condensed atom and an atom in another mode is twice that between two condensed atoms. Hence one might expect the presence of thermal uncondensed atoms to shorten  the collapse time compared to the GP prediction, as observed.  Furthermore, this might be approximately corrected for by using an increased magnitude effective scattering length in the GP model. However the uncondensed fraction is  much less than 10\% of the total number of atoms \cite{Claussen}, so it seems unlikely that this effect is large enough to account for the discrepancy. Furthermore, Roberts \textit{et al.}  \cite{Robertscollapse} reported that  the critical number for collapse $N_{\rm{cr}}$ was insensitive to varying the temperature. Therefore we do not expect finite temperature extensions of the GP theory to explain the discrepancy \cite{finite temperature}.

Another possibile origin is quantised atom field effects. These might arise due a breakdown of the factorisation assumption Eq.(\ref{factorisation}). There have been several suggestions for how the quantised field might influence the collapse \cite{UedaLegget,DuineStoof,Yurovsky}. Furthermore, Nozieres \cite{Nozieres} has emphasised that only for positive scattering lengths does an energy barrier protect BECs from fragmentation into many populated states. For negative scattering lengths, mean-field energy is released when atoms scatter from the condensate into other modes. Fragmentation could increase the effective scattering length by up to a factor of two.

In order to investigate the behaviour of a fully quantised atom field, we have used the gauge-P function approach recently developed by Deuar and Drummond \cite{DeuarDrummond}. This method overcomes some of the problems that plague stochastic simulations based on the positive P-function quasi-probability distribution \cite{Drummond,Steel}. We were computationally limited to simulations in one spatial dimension and found agreement with the GP collapse times at the one percent level.  

Although this preliminary work does not provide evidence for quantum field effects, it is important to extend the fully quantised field modelling to three spatial dimensions, and hence to use actual experimental parameters. As shown in Fig.~1, there are parameters for which the GP theory does agree with experiments. One approach is the recently developed perturbation theory which extends the GP model to include normal and anomalous densities of the quantum field \cite{HollandParkWalser}. This method has recently been successfully applied to the formation of atom-molecule condensates in $^{85}$Rb \cite{KokkelmansHolland}.


\begin{acknowledgments}
This research was supported by the Australian Research Council and by an award under the Merit Allocation Scheme on the National Facility of the Australian Partnership for Advanced Computing \cite{APAC}. 
\end{acknowledgments}


\begin{thebibliography}{99}
\bibitem{3dnumerics} These numerical solutions were performed on spatial grids as large as 512x256x256.
\end{thebibliography}
\end{document}